# IMPLEMENTING A GREEDY CHAIN ROUTING TECHNIQUE WITH SPREAD SPECTRUM ON GRID-BASED WSNS

Hossein Sharifi Noghabi [1], Arash Ghazi askar [1], Arash Boustani[2] and Arash Moghani[1], Motahareh Bahrami Zanjani[2]

[1]Department of Computer Engineering and Information Technology,
Sadjad Institute of Higher Education, IRAN
[2]Department of electrical Engineering, Wichita State University, USA
(H.sharifi219/ a.ghazi2003/ a.moghani206)@sadjad.ac.ir, (axboustani/ mxbaharamizanjani@wichita.edu)

## ABSTRACT

*Wireless Sensor Networks (WSN) are set of energy-limited sensors, which recently have been point of interest due to their vast applications. One of the efficient ways to consume energy in these networks is to utilize optimal routing protocols. In this approach, we proposed a greedy hierarchical chain-based routing method, named, PGC (stands for Persian Greedy Chain) which route the network applying Spread Spectrum codes as a mask given to the grid cells. Due to similarities between the proposed method in this article and LEACH protocol, we compare this routing protocol with the proposed model from diverse aspects in the simulation section such as remaining energy and being fault tolerant and reliable. The results prove that presented method is more robust and efficient.*

## 1. INTRODUCTION

In recent years, Wireless Sensor Networks had been developed in many fields like medical science, military and even commercial aspects.one of the critical concerns of scientists in WSN, is optimal consuming of energy. Since sensor's batteries are not rechargeable or charging them is too costly, we should apply some special methods to minimize the energy consumption .Obviously, with reducing the energy consumption, the lifetime of the network will be maximized.

One of the solutions is to utilize an optimal routing in WSN. From structure designing view, routing techniques are divided into three parts: Flat, Hierarchical and Location-based.

In flat routing protocols, all of the nodes play same role in aggregating and sending data and routing process. In location-based routing algorithms, sensors are controlled by places they hold. In hierarchical routing algorithms nodes are hierarchical structured and network is divided into particular regions called Cluster [1].

For hierarchical structures, LEACH algorithm is discussed [2]. LEACH includes a distributed arrangement of clusters in which the Cluster Head (CH) is responsible for data aggregation and data transmission to the sink.

DOI : 10.5121/ijwmn.2012.4414    195

Another solution which leads to reducing energy consumption and consequently increasing the network lifetime in WSN, is taking advantage of Grid-cells.

In [3], it is illustrated that applying grid cells in WSN significantly reduces energy consumption. In each grid only one Gateway (GW) node should be active and sense the environment, hence, all other nodes will be slept. When the energy if the GW reaches to a minimum threshold, the GW is able to substitute another node, so the explained process will be repeated all over again. By utilizing GAF protocol; it is highly likely to save 40 - 60% of energy.

In [4] a new and secure method for data aggregation is introduced. This method is based on Grid cells, and one single mask is considered for a group of sensors, using PCC (Persian Chip Code).

This manner causes each group of sensors to be addressed uniquely and data to be delivered would be encoded. It is clear that encoding each data leads to more confidentiality and security in the network [4, 5].

Grid shape is also effective in optimizing the energy consumption of whole WSN. Results had shown that hexagonal grids due to full coverage, will improve network functionality in consuming energy, end - to - end delay, fault tolerance and security [6]. In the other types of networks dividing the environment will also increase the utilization [7].

## 2. PREVIOUS WORKS

### 2.1. Hierarchical algorithms in organizing network structure

LEACH algorithm is a cluster-based algorithm and clustering the nodes is based on receiving signal strength. The responsibility of transmitting data to the sink is with cluster heads.

In LEACH protocol, only 5 % of all nodes take the chance of being cluster head. All data processing, like data aggregation and propagation, is carried out locally in every cluster. LEACH algorithm has its own disadvantages too. Each time the algorithm is run in the network, the lifetime of sensors in the whole network will be decreased. Moreover, the energy of CH nodes is consumed more quickly than the other nodes [8]. In LEACH, data is transmitted directly from nodes to cluster heads, as a result, if there is a long distance between one of the nodes and the cluster head, the non-cluster head node should consume too much energy for delivering the information to the cluster head. This problem is strongly against LEACH, because it consumes a great deal of energy to transmit data to the sink. Nevertheless, the transmitting data might be corrupted or lost in the path and make the energy consumption much higher. LEACH also has a large overhead in producing tree among the nodes [1].

### 2.2. Grid-cell production algorithms

As mentioned earlier, one node can be active in each grid and alter other nodes to sleep mode. The activated node is often called Coordinator or Gateway node. GAF and SPAN are two methods for creating grids.

In GAF, the environment is divided into virtual cells and determinations about which node is sleep and which node is GW are made based on system information. The only problem in this method is that geographical locations of sensors are needed. In SPAN, each node can determine whether being coordinator or not. Coordinator election is performed periodically; just to prevent high energy consumption in GW nodes. One of the worst disadvantages of applying SPAN is electing too much coordinator, whereas in this method the total number of coordinator should be minimal [3].

### 2.3. Persian Chip Code

The proposed method in PCC, use a recursive algorithm and produces Orthogonal Chip Codes on a matrix in order to work on the CDMA-based networks which have the Auto Correlation, Cross Correlation and Hamming Property features. The below matrix is used for creating PCC:

$$P_{4^n} = \begin{bmatrix} P_{4^{n-1}} & P_{4^{n-1}} & P_{4^{n-1}} & \sim\mathbf{P_{4^{n-1}}} \\ \sim P_{4^{n-1}} & \mathbf{P_{4^{n-1}}} & \sim P_{4^{n-1}} & \sim P_{4^{n-1}} \\ P_{4^{n-1}} & P_{4^{n-1}} & \sim \mathbf{P_{4^{n-1}}} & P_{4^{n-1}} \\ \mathbf{P_{4^{n-1}}} & \sim P_{4^{n-1}} & \sim P_{4^{n-1}} & \sim P_{4^{n-1}} \end{bmatrix}$$

$$\forall n \in \mathbb{N} \; ; \; P_{4^0} = [0]$$

### 2.4. Energy aware and highly Secured Data Aggregation for Grid-based Asynchronous Wireless Sensor Networks

One of the related methods is to encode and aggregate data using PCC (Persian Chip Code). For data aggregation, the cluster heads transmit the encrypted data to the sink and the sink will decrypt it in order.at first, the sink divides the network into grid-cells, then it specify a set of PCC codes to the sensors within each grid for encoding data.as a result, the network will be addressed just like a non-WSN network with IP and Subnet [4].

## 3. PROPOSED MODEL

In the proposed model (PGC), a method for optimal consuming of energy and prolonging the lifetime of WSN is provided. This method will improve energy consumption in the network by making the best use of a greedy routing algorithm and smart movements of sinks in the region of interest. In this section we will explain different phases of proposed model.

### 3.1. Proposed Routing algorithm

After forming our hexagonal grid cells and defining the active node in each grid, the discussed algorithm in 2.4 Section will be run on all grids. After running PCC on whole network, a hierarchical address similar to IP is assigned to all of the GWs and each GW, according to its mask, will communicate with other GCs.

At first, Sink broadcasts an initial HELLO message. Then, all of the receiving messages are examined, and the nodes which their message were received with maximum transmission power will be selected as "immediate GWs". These "immediate GWs" are the nearest GWs to the sink .We introduce a parameter, namely "Next_Hop_Sink" (NHS). This parameter is given to the nearest adjacent GWs which replied the initial message with their minimum transmission power (in the other word, sink received with maximum transmission power). GWs holding NHS parameter , transmit the data  directly , immediately  and  without any routing to the sink .Main idea in this model is that each GW transmit data to the adjacent GW with least Mask number - which was produced and assigned to each grid earlier, by PCC. This numbering method indicates that other GCs (which are non-immediate) which are nearer to the sink, would obtain the smaller mask number in comparison to the farther GCs. for example if the sink is located at the right side of the network, then the GCs located at the right side of the GCs containing "immediate" GWs, have smaller mask number. Note that similar to ESTOC [4], in this method the sink eventually has all of the data from all of the sensors in the network.

In addition to every sensor node, each GC has its own energy parameter (denoted by G for Green, Y for Yellow and R for Red) for whole GC. Green GCs are grid cells which have at least one Green node within. While all of the nodes in the GC become yellow, the GC state becomes Yellow too. It is obvious that if all the nodes become red, the GC state turns to Red. Red GCs would not participate in routing process. The priority is with Green GCs. in the absence of Green GCs in the predefined route (which were calculated earlier e

The last important thing to note is that if a GW attempts to send data and due to energy consumption, there is no way reachable to sink by its neighbor grid cells, the GW would send data directly to sink. Of course, if any available GC is present between the dead GCs and sink, data would be delivered to those GCs first.

### 3.2. Coding and decoding manner in proposed model

In chain-based routing, every time the data is encoded and decoded, large overhead will be imposed to the network. The solution is that sink broadcasts all the keys, at the very first moment to all of the grids. By doing this, each GW stores mapping of all grids to their corresponding keys, so encoding and decoding will be done easily. After changing the GW and electing a new one, the grid keys will be transferred from old GW to the new one.

### 3.3. Distributing energy consumption by moving sinks and sliding windows

To achieve the goal of distributing energy consumption, multiple sinks are used asynchronously. Whenever the average energy of GWs adjacent to the sink reach to a particular threshold, in order to prevent centralized energy consumption, another sink can be activated. It is clear that the addressing process should be start again and new "immediate" GWs will be chosen.

Now suppose that all of the sinks had been selected one time and it's the first sink's turn .if the old low-energy GWs continue being as GW, and the other GWs never have a chance to act as the "immediate" GWs, the overall lifetime of the network would be decreased. thus in proposed method , when the energy of GWs in the window reach to a particular threshold , the GWs send a message to sink and sink would consequently omit the mask address of the low-energy GW. Then sink would move to the other side of the grid and the process of choosing the "immediate" GWs would be started again. There's no need to mention that one sink should be active at a time.

### 3.4. Examining an instance

In order to clarify the proposed model, an instance illustrated to plot the routing technique. As it is shown in Figure.1, the proposed routing technique implies that routing through the network from source to sink should be performed in a greedy manner. First step is distributing codes generated by PCC to all grid-cells. In the example, the chip codes simplified to simple decimal numbers, thus, each grid now has a specific unique ID. Assume that GW in grid-cell 25 attempts to transmit data. The GW would send data to its least-numbered adjacent GW (here, active GW of grid-cell 19). So data would be delivered similarly to the next GW, until it reaches the grids holding NHS parameter. In this phase data would be delivered directly to the sink, located in top right of the network.

After some time which the average energy of our three grid-cells holding NHS parameter reaches to a particular threshold, the sink would move to the left side and all the process start all over again.

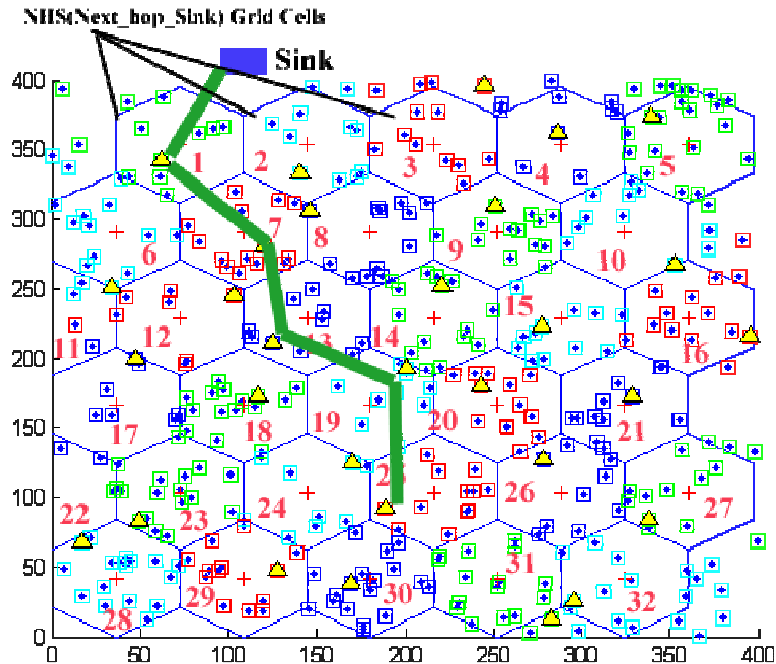

Figure 1. Routing data sourced from GC 26 to sink, passing through GCs 19 – 13 – 7 – 1 .The sliding window contains the grid numbers which hold NHS parameter.

## 4. NETWORK MODEL

According to previous section, our proposed approach has two main advantage in data gathering in comparison to LEACH : first , in the case that data transmission is not back-to-back stream of data and data is transmitted out- of-order and with an exponential distribution , our proposed model has better performance. In periodic transmission by all of the nodes, LEACH can obtain similar performance to the proposed method, while in other case; the proposed method has better performance. In order to prove the results, extensive simulations performed with MATHLAB 2008.

The second advantage of the proposed method is that due to using PCC in proposed method, data transmission in noisy environments is much better than LEACH, and data corruption has much lower affection to the network which uses PGC, in comparison to LEACH. This will cause reducing retransmissions in network which will leads to more energy saving of nodes and increase lifetime of entire network. Remaining energy of sensors in proposed model and LEACH per 35000 packets in network is illustrated in Figure 1. Due to Figure 1, in equal times of algorithm execution, proposed model has consumed less energy than LEACH. As depicted in Figure 2, available duty cycle of our proposed model and LEACH, each of them in two cases of 99.0 % and 99.5 % packet loss is plotted. Due to using PCC in our proposed model, fault tolerance in our method is better. In each two cases, even by applying noise, receiver can decode information and obtain data.

Table 1 illustrates required information for simulation and amount of environment changes in different cases of network operations.

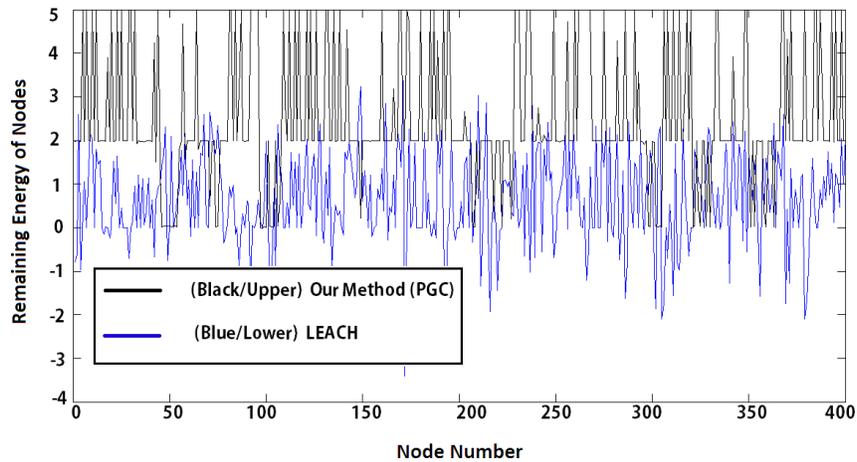

Figure 2. Remaining Energy of nodes after transmitting 35000 packets. The upper wave (depicted in black) which is our method (PGC) has more energy than nodes running LEACH protocol (which is depicted in blue, lower wave).

Table 1. Main Simulation Parameters

| Parameter | Value |
| --- | --- |
| Network size | 400*400 meter |
| TX Power / RX Power | 15mW/10m / 13 mW/10m |
| Processing Power | 3 mW/10m |
| Initial Energy per Node | 5J |
| Minimum threshold of energy | 0.03 J |
| Number of Nodes | 400 sensors |
| ε fs | 10 PJ/bit/m4 |
| Carrier frequency | 2.4 - 2.48 Ghz (Zigbee) |
| Average of sensor per grid-cell | 13 sensors |
| Number of grid-cell | 30 grid-cell |
| Grid-cell structure | HCn6 |
| Virtual grid Creator Algorithm | GAF Algorithm |
| Chip Code Size | 32 bit- Orthogonal Asynchronous- Chip Code |
| Orthogonal Chip Code creator | Persian Algorithm Chip Code |
| Communication Multiplexing | CDMA - FHSS |
| Noise value | 99.0 & 99.5 % packet loss |

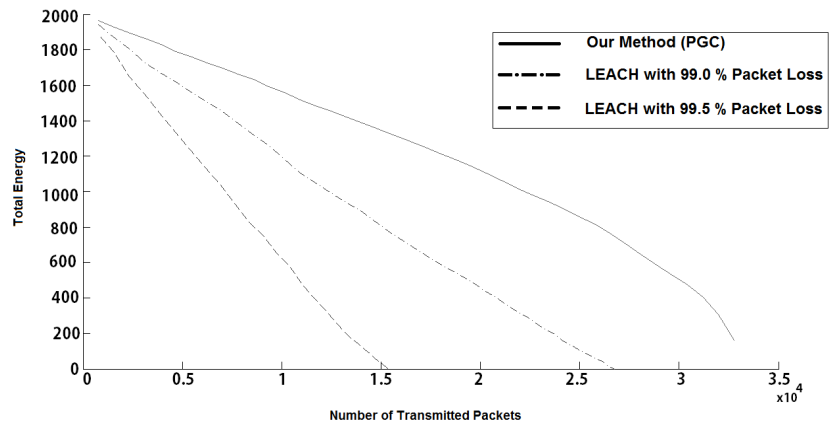

Figure 3. Evaluating two protocols in noisy environment. The upper line is our proposed model which has least packet loss per 35000 packets, the middle dotted-dashed line is LEACH with 99.0 % packet loss and the lower dashed line is LEACH with 99.5 % packet loss.

## 5. CONCLUSION

Obviously , energy consumption is one of the important challenges in WSN and by using an appropriate routing method , energy consumption in these networks would be enhanced and lifetime of sensors and whole network will be increased.in this paper, we provided an optimal chain-based routing method in which , sensors in a grid-based WSN choose the best path to aggregate data and transmit all of the data to the sink.in the proposed method , special Orthogonal chip codes in spread spectrum  are used for addressing each GC and routing the network so that each GW in each GC , select the GC with least mask number for transferring data through it .this selection will cause reducing energy consumption significantly . Additionally, existence of multiple sinks which are activated asynchronously, and a sliding window in sink's movements, will optimize the energy consumption which consequently prolong lifetime of sensors and WSN.